\documentclass
[
    12pt,				
	english,            
	listof=totoc,		
	bibliography=totoc, 
	parskip,
]{scrartcl}

\usepackage{geometry}
\usepackage[english]{babel}
\usepackage{amsmath}
\usepackage{amsfonts}
\usepackage{amssymb}
\usepackage{booktabs}
\usepackage{longtable}
\usepackage{xurl}
\usepackage{graphicx}
\usepackage[table]{xcolor}
\usepackage{libertinus} 
\usepackage[printonlyused]{acronym}
\usepackage{varioref}
\usepackage[colorlinks=true, allcolors=blue!50!black]{hyperref}
\usepackage{csquotes}
\usepackage{multicol}
\usepackage{array}
\usepackage{comment}

\usepackage{libertinus}
\usepackage{lmodern}
\usepackage{listings}
\lstdefinelanguage{JavaScript}{
  keywords={typeof, new, true, false, catch, function, return, null, catch, switch, var, if, in, while, do, else, case, break, const, let, async, export},
  ndkeywords={class, export, boolean, throw, implements, import, this},
  identifierstyle=\color{black},
  sensitive=false,
  comment=[l]{//},
  morecomment=[s]{/*}{*/},
  morestring=[b]',
  morestring=[b]",
  morestring=[b]`
}

\setuptoc{toc}{totoc}
\usepackage[headsepline]{scrlayer-scrpage}	
\clearpairofpagestyles		            
\automark{section}			            
\ohead{\normalfont\headmark}			
\ofoot{\normalfont\thepage}			    

\usepackage{biblatex}
\bibliography{literature}

\usepackage[capitalise,noabbrev,nameinlink]{cleveref}

\begin{document}



\begin{titlepage}
    \begin{minipage}[c]{0.45\textwidth}
    \includegraphics[width=0.625\textwidth]{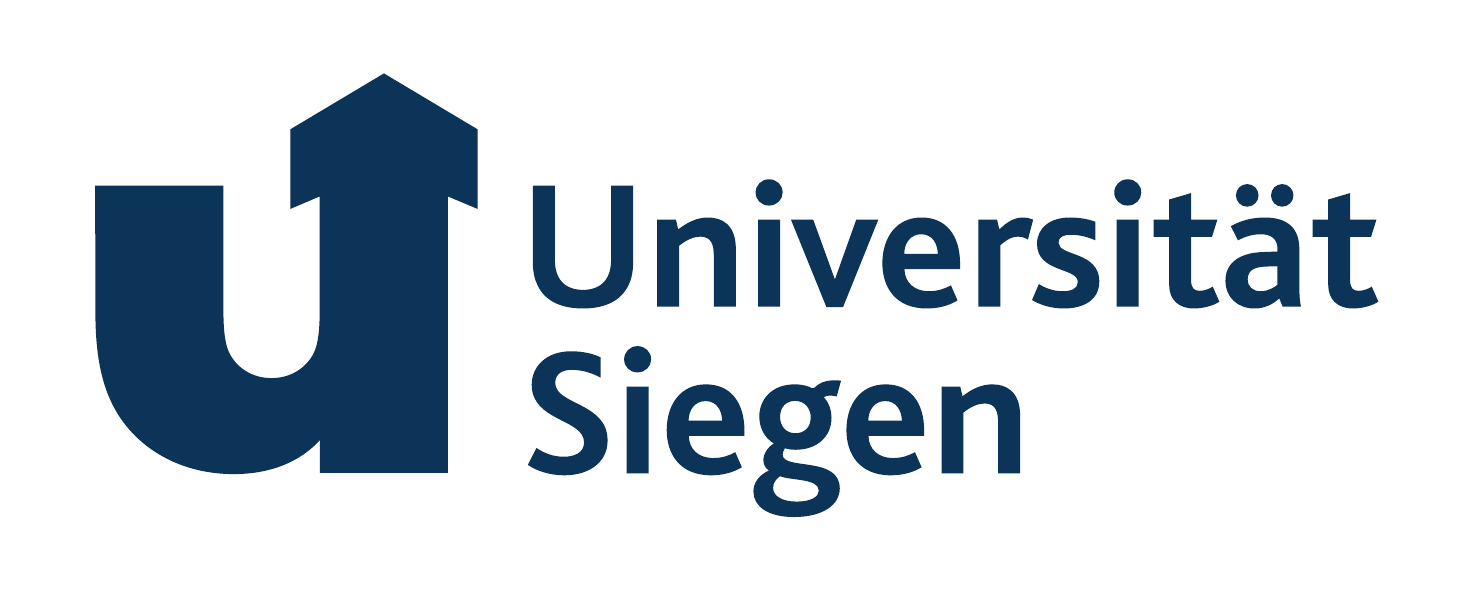}
    \end{minipage}\hfill
    \begin{minipage}[c]{0.45\textwidth}
    \includegraphics[width=1\textwidth]{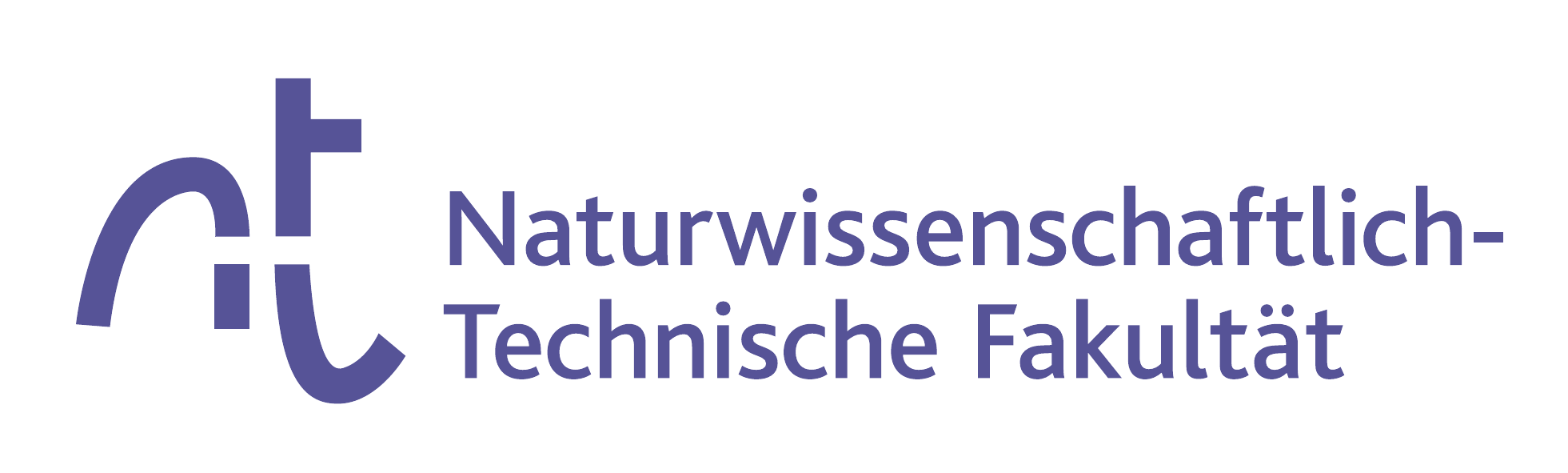}
    \end{minipage}	
	\vspace*{1cm}
    \begin{center}
        \LARGE
        Project Report\\
        \vspace*{1cm}
        \huge
        Informed Dataset Selection
    \end{center}
    \vfill
    \normalsize
    \begin{tabular}{ll}
        Authors & Abdullah Abbas \\
                & Michael Heep \\
                & Theodor Sperle \\
        \\
        Submitted & 28 August 2025
    \end{tabular}

\addsec{Abstract}

The selection of datasets in recommender systems research lacks a systematic methodology. Researchers often select datasets based on popularity rather than empirical suitability.
We developed the APS Explorer, a web application that implements the Algorithm Performance Space (APS) framework for informed dataset selection. 
The system analyzes 96 datasets using 28 algorithms across three metrics (nDCG, Hit Ratio, Recall) at five K-values. We extend the APS framework with a statistical based classification system that categorizes datasets into five difficulty levels based on quintiles. We also introduce a variance-normalized distance metric based on Mahalanobis distance to measure similarity.
The APS Explorer was successfully developed with three interactive modules for visualizing algorithm performance, direct comparing algorithms, and analyzing dataset metadata.
This tool shifts the process of selecting datasets from intuition-based to evidence-based practices, and it is publicly available at \url{datasets.recommender-systems.com}. 
\end{titlepage}

\newpage
\tableofcontents

\section{Introduction}
\label{sec:introduction}

The selection of datasets plays a critical role in shaping the validity, reproducibility, and generalizability of research in recommender systems \cite{Beel24, cremonesi2021crisis, schnabel2022we, beel2025green}. Yet, despite its importance, the selection of datasets in recommender systems is often unclear rather than a systematic evaluation of their suitability and the analyses of publications at major venues such as ACM RecSys reveal that researchers frequently rely on a small set of benchmark datasets without providing substantive justification beyond public availability or precedent \cite{beel2019pruning,Beel24,cremonesi2021crisis, wegmeth2024emers}. While such benchmarks have facilitated comparability, their dominance could limit experimental diversity and may overstate algorithmic progress on narrowly defined problems \cite{cremonesi2021crisis,dacrema2019neural}.

Previous research has shown that intrinsic dataset characteristics, such as sparsity or scale, can influence algorithmic performance\cite{adomavicius2012impact}. However, these factors alone often fail to fully explain performance variations \cite{adomavicius2012impact,Chin22}. To address these shortcomings, Beel et al. \cite{Beel24} introduced the \acf{APS} framework, which represents datasets according to the empirical performance profiles of multiple algorithms. In this representation, inter-dataset distances reflect how similarly algorithms behave, offering a basis for identifying datasets that are redundant, diverse, \enquote{solved}, or \enquote{challenging} based on performance.

Building on this \ac{APS} framework foundation, our work introduces the \ac{APS} Explorer. The \ac{APS} Explorer extends the \ac{APS} framework by integrating algorithm performance analysis, statistical classification, and interactive visualization. This allows \ac{APS} Explorer to address a long-standing gap in recommender systems research: the lack of a transparent, data-driven methodology for dataset selection. In doing so, it contributes to more reproducible experimentation, more representative benchmarking, and ultimately, a more rigorous understanding of algorithm generalizability in diverse recommendation scenarios.

Furthermore, the \ac{APS} Explorer incorporates several methodological enhancements to assess key factors such as dataset difficulty, as well as the similarities and dissimilarities between datasets. These methodologies include a quintile-based statistical classification system for dynamic dataset difficulty assessment \cite{Agresti18,Tukey77}. It also incorporates a variance-normalized similarity metric based on the Mahalanobis distance \cite{Mahalanobis36}. This is combined with an exponential decay transformation \cite{Scholkopf02} for more robust similarity estimation. Performance results from 29 algorithms on 96 datasets were considered, evaluated with metrics such as \ac{nDCG}, \ac{HR} and Recall and assessed across multiple cut-off values. Using this data, the \ac{APS} Explorer offers new capabilities to researchers, like the visualization of dataset diversity. It also supports comparison of algorithmic behavior and it helps examine metadata-driven risk factors such as user–item imbalance and interaction sparsity.

In addition, the project resulted in a research paper, which has been accepted for presentation at the ACM Conference RecSys 2025 \cite{recsys25}.

The remainder of this paper is organized as follows: \cref{sec:relatedWork} reviews existing approaches to dataset selection and positioning in recommender systems research. \cref{sec:background} provides the theoretical foundation of the \ac{APS} framework and its mathematical background. \cref{sec:mainPart} presents our methodology for the selection of datasets and the empirical analysis of algorithmic performance patterns. \cref{sec:implementation} details the technical architecture and features of the \ac{APS} Explorer web application. Finally, \cref{sec:conclusion} summarizes our contributions and discusses implications for future recommender systems research.

\section{Related Work}
\label{sec:relatedWork}

Dataset selection in recommender systems has increasingly come under scrutiny due to its significant influence on experimental outcomes \cite{adomavicius2012impact,Beel24,Chin22, 10.1145/3705328.3759342}. Despite its importance, most prior work has either overlooked this issue or addressed it only superficially. For instance, Beel et al. \cite{Beel24} found that in RecSys 2023, 44\% of papers did not offer justification for the selection of dataset. Similarly, Chin et al. \cite{Chin22} highlighted that dataset selection is often arbitrary and rarely related to experimental objectives. Datasets like MovieLens and Amazon dominate empirical studies \cite{Beel24}, yet this ubiquity is rarely accompanied by a rigorous analysis of their suitability for specific tasks or algorithmic properties.

Early work on understanding the impact of dataset properties focused on how intrinsic characteristics such as sparsity and interaction density affect the performance of recommendation algorithms Chin et al. \cite{Chin22} clustered 51 datasets based on statistical features and demonstrated that algorithms like RP3beta and WMF exhibit significantly different behaviors across clusters. Their results showed up to 45\% performance variation depending on the sparsity and scale of the dataset. However, they also noted inconsistencies: datasets with similar structural profiles sometimes produced divergent algorithmic outcomes, suggesting that surface-level metadata may not capture all factors influencing performance.

To address these limitations, Beel et al. \cite{Beel24} extended the concept of \acf{APS}, originally introduced by Tyrrell et al. \cite{tyrrell2020personas}, to the domain of recommender systems. In their adaptation, \ac{APS} represents datasets as vectors in a high-dimensional space, where each dimension corresponds to the performance of a specific algorithm. In this space, dataset distances reflect empirical diversity: datasets that produce similar algorithm performance profiles are placed close together, while those inducing different outcomes appear further apart. This shift from feature-based to outcome-based representations offers a more reliable and reproducible mechanism for identifying diverse or redundant datasets.

 Beel et al. \cite{Beel24} conducted on 29 algorithms in 95 datasets, revealed that many commonly used datasets, particularly Amazon subsets, are tightly clustered in \ac{APS}, indicating similar algorithmic behavior and low diversity. In contrast, datasets such as Docear or FilmTrust were positioned further from this cluster, suggesting their potential to introduce novel evaluation challenges. Beel et al. also identified \enquote{solved} datasets, i.e., most algorithms perform well, and \enquote{challenging} ones, i.e., all algorithms perform poorly. This offers a new way to prioritize datasets for future benchmarking.

Many concerns have previously been raised regarding the reliance on traditional benchmarks. Cremonesi and Jannach \cite{cremonesi2021crisis} and Ferrari Dacrema et al. \cite{dacrema2019neural} criticized the field of recommender systems for relying on a narrow set of datasets without questioning their validity. They argued that many advances in algorithm design yield marginal gains on these benchmarks, potentially inflating progress while reducing external validity. Furthermore, studies such as Adomavicius and Zhang \cite{adomavicius2012impact} emphasized the unpredictable relationship between data characteristics and algorithmic results.

Although performance-based dataset representations like \ac{APS} offer a convincing alternative, they do not necessarily invalidate the role of metadata. Several researchers have proposed hybrid strategies that combine the properties of intrinsic datasets with performance metrics to support algorithm selection or meta-learning \cite{tyrrell2020personas,fan2023movielens, vente2022greedy, wegmeth2024recommender}. However, a key challenge remains: how to systematically capture the aspects of a dataset that influence algorithm behavior, particularly in scenarios where metadata fails to predict outcomes.

In summary, the contribution of the presented study differs from previous studies in several key aspects. Although earlier studies introduced the \ac{APS} framework and demonstrated its potential to analyze dataset diversity \cite{Beel24,tyrrell2020personas}, this work extends the framework with a quintile-based statistical classification system for dataset difficulty \cite{Agresti18,Tukey77} and a variance-normalized similarity metric based on Mahalanobis distance \cite{Mahalanobis36,schnabel2022we}. In addition, the concepts are realized through the \ac{APS} Explorer, a publicly available web application that integrates algorithm performance analysis, dataset comparison, and metadata-based risk assessment into a unified platform. Through these methodological extensions and practical implementation, the presented approach establishes a novel approach to evidence-based dataset selection in recommender systems research.

\section{Background}
\label{sec:background}

This section provides the necessary theoretical and methodological background to understand the proposed solution, the Algorithm Performance Explorer, a web application that implements the \acf{APS} framework for informed dataset selection. \Cref{subsec:recsys} introduces recommender systems, with particular attention to performance metrics, k-values, and risks such as bias and cold starts. \Cref{subsec:backgroundAPS} presents the \ac{APS} framework, which serves as the basis for our approach which uses a quintile-based statistical classification system for dataset difficulty and a variance-normalized similarity metric to support evidence-based dataset selection. Finally, \cref{subsec:methodology} outlines the mathematical methods applied in this work, including \ac{PCA}.

\subsection{Recommender Systems}

\label{subsec:recsys}
This subsection introduces background knowledge on recommender systems, covering dataset structure, evaluation metrics, and common challenges such as data sparsity, bias, data splitting\cite{baumgart2024fold}, and cold-start\cite{beel2025green} risk.

An important feature for a good recommender system is its data. A dataset for a recommender system typically consists of the following features, displayed as a table: users, items and interactions.

\begin{table}[h!]
\centering
\caption{Example of user-item interactions}
\label{tab:user_item}
\begin{tabular}{cccc}
\toprule
User & Item 1 & Item 2 & Item 3 \\
\midrule
U1 & 3 & 5 & 2 \\
U2 & 1 & 3 & 5 \\
U3 & 4 & 2 & 4 \\
U4 & 2 & 4 & 2 \\
\bottomrule
\end{tabular}
\end{table}

Table \ref{tab:user_item} shows an example of a dataset for a recommender system. The table consists of a set of Users $U$, a set of Items $I$, and a set of Ratings $R$. A user is an individual who interacts with the system and receives personalized recommendations. An item is a product, service, or piece of content that can be recommended to users. An interaction represents an action a user takes with an item, such as viewing, rating, or purchasing it.

In recommender systems, datasets can contain different types of instances, depending on the data collected. Implicit and explicit datasets are distinguished. An explicit dataset contains data where users explicitly rate items, for example, a rating from 1 to 5. An implicit dataset consists of implicit user feedback, such as clicks, views, purchases, or other types of interactions, rather than on explicit ratings.

The goal of a recommender system is to suggest items to users that match their preferences and interests. One common technique used for this is collaborative filtering. In collaborative filtering, a user is compared to other users with similar interaction patterns based on shared items. A widely used approach to implement collaborative filtering is matrix factorization. In matrix factorization, the user–item interaction matrix is decomposed into latent factors that capture hidden relationships between users and items. These latent representations help the system predict how likely a user is to interact with items, for which a user has no interactions with. One of the main challenges of collaborative filtering is dealing with new users or new items that have no prior interaction history. Since the system relies on user-item interactions to generate recommendations, it cannot make meaningful suggestions for new users or items. This issue is known as the cold-start problem.

Another common challenge of recommender systems and collaborative datasets is data sparsity -- a phe\-nom\-e\-non where the interaction matrix contains many missing (empty) values. One reason for data sparsity is, that users only interact with a fraction of items. 

\subsubsection{Performance Metrics}
\label{subsec:metrics}

Now let's take a look at some common metrics used in recommender systems. In general, there are many metrics in recommender systems that are also found in traditional machine learning. These metrics can differ depending on the goal of the recommender system, the specific task it is designed to perform, and the nature of the dataset being used. 

One metric for evaluating recommender systems is \acf{nDCG}. \ac{nDCG}@N evaluates the quality of recommendations by checking whether more relevant items appear higher in the ranking. The equation for \ac{nDCG}@N is described as:

\begin{equation*}
    \text{nDCG}_p = \frac{\text{DCG}_p}{\text{IDCG}_p}
    = \frac{rel_1 + \sum_{i=2}^{p} \frac{rel_i}{\log_2(i + 1)}}{\text{IDCG}_p}
\end{equation*}

\begin{equation*}
    \text{IDCG}_p = rel_1^{*} + \sum_{i=2}^{p} \frac{rel_i^{*}}{\log_2(i + 1)}
\end{equation*}

where:
\begin{itemize}
    \item$p$: Number of top-ranked items considered (cut-off rank)
    \item$rel_i$: Relevance score of the item at position $i$ in the ranked list
    \item$\text{DCG}_p$: Discounted Cumulative Gain up to position $p$
    \item$\text{IDCG}_p$: Ideal DCG up to position $p$ in the ranked list
    \item$\log_2(i + 1)$: Logarithmic discount factor that reduces the contribution of items at lower ranks
\end{itemize}

The DCG is calculated using the relevance scores of the recommended items, where lower-ranked items contribute less due to the logarithmic discount.
The IDCG uses the ideal relevance scores $rel_i$, sorted in the best possible order to represent a perfect recommendation.
The final \ac{nDCG} score ranges from 0 to 1 and reflects how close the actual ranking is to the ideal one.

For each metric discussed in this subsection, a corresponding k-value can be defined. The k-value determines how many of the top-ranked items are taken into account when computing the metric. For example, if the k-value in \ac{nDCG} is set to 10, only the top 10 recommended items are evaluated.

Another metric, which evaluates the ranking of a recommender system is the so called \acf{HR}. The formula for \ac{HR} is described as:

\begin{equation*}
    \text{Hit}_{u,N} = 
    \begin{cases}
    1, & \text{if at least one relevant item is among the top-}N\text{ recommendations} \\
    0, & \text{otherwise}
    \end{cases}
\end{equation*}

\begin{equation*}
    \text{HR@N} = \frac{1}{|U|} \sum_{u \in U} (\text{Hit}_{u,N})
\end{equation*}

\Acf{HR} evaluates whether at least one relevant item is present in the top‑N recommended items for a user. For a single user, the value is 1 if any of the recommended items are relevant, and 0 otherwise. The overall \ac{HR} is calculated as the average across all users. A higher \acl{HR} indicates that the recommender system is effective at placing relevant items within the top‑N positions of its recommendations.

Another important metric for evaluating recommender systems is Recall@N. It measures the proportion of all relevant items that are successfully retrieved within the top‑N recommended items. The formula for Recall@N is given by:

\begin{equation*}
\text{Recall@N} = \frac{|\text{Recommended}_N \cap \text{Relevant}|}{|\text{Relevant}|}
\end{equation*}

Recall@N evaluates how many of the relevant items for a user appear in the top‑N recommendations. A higher recall value indicates that the recommender system is effective at retrieving a large portion of relevant items. Unlike \acl{HR}, which only checks for the presence of at least one relevant item, Recall takes into account all relevant items and rewards systems that retrieve as many as possible.

\subsection{Algorithm Performance Space}
\label{subsec:backgroundAPS}

\acfp{APS} offer a structured framework for selection of datasets for recommender systems by mapping datasets into a multidimensional space defined by algorithmic performance. The core claim here is that \ac{APS} enables more informed and reproducible decisions about dataset selection by visualizing and quantifying diversity in terms of how algorithms perform, rather than relying on intrinsic dataset properties such as sparsity or domain. This reframing of the dataset selection problem is motivated by the observation that traditional justifications such as popularity, availability or assumed real-world relevance, lack rigor and may misrepresent the generalizability of algorithmic performance. \cite{Beel24}

The \ac{APS} is constructed as an $n$-dimensional space, where each dimension corresponds to the performance of one specific algorithm. Each dataset becomes a point or vector in this space, characterized by its performance profile across these algorithms. The Euclidean distance between any two points reflects how similarly or differently algorithms perform on the corresponding datasets. When two datasets are close in this space, it implies that all tested algorithms perform in a consistent fashion across them. This consistency suggests that a new algorithm is also likely to behave similarly across those datasets, thus potentially reducing the need to evaluate it on all datasets within that cluster.

This mechanism directly addresses a key problem in experimental design: redundancy. If several datasets yield nearly identical algorithmic performance profiles, then they contribute little additional information about the behavior or generalizability of a new algorithm. Instead, by focusing on datasets that are widely separated in \ac{APS} (for example, diverse in performance characteristics), researchers can evaluate algorithmic robustness under more varied conditions. In essence, distance in the \ac{APS} is a proxy for the experimental value of a dataset in the context of algorithm comparison.

\begin{figure}
    \centering
    \includegraphics[width=.6\textwidth]{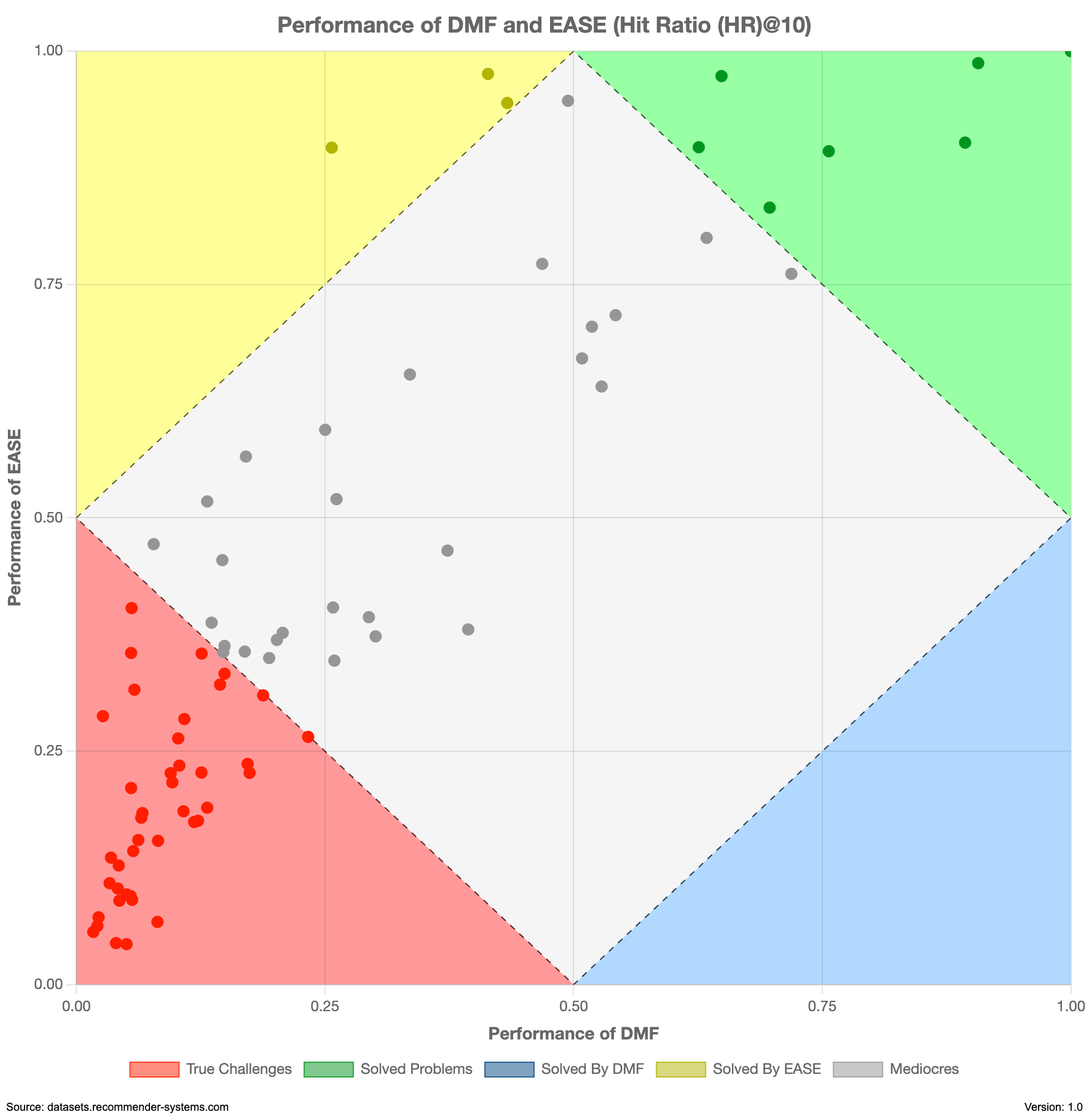}
    \caption[Example of comparing two algorithms]{Example of comparing DMF and EASE algorithms with \acs{HR}@10}
    \label{fig:algorithmComparison}
\end{figure}

To illustrate, \cref{fig:algorithmComparison} shows a simplified two-dimensional \ac{APS} involving only two algorithms, A1 and A2. Each dataset is plotted as a point where the x-coordinate represents the \ac{HR} score of A1 and the y-coordinate the \ac{HR} of A2. If a dataset lies near the top right corner, it indicates high performance for both algorithms, suggesting the underlying recommendation problem is relatively easy or solved. Datasets near the bottom left corner show poor performance across the board, indicating complex or unresolved problems. Datasets that fall elsewhere, such as those where A1 performs well but A2 poorly, help reveal algorithm-specific strengths and weaknesses. Thus, the spatial arrangement of datasets within \ac{APS} becomes a lens through which to assess problem difficulty and algorithm capabilities.

\begin{figure}
    \centering
    \includegraphics[width=.6\textwidth]{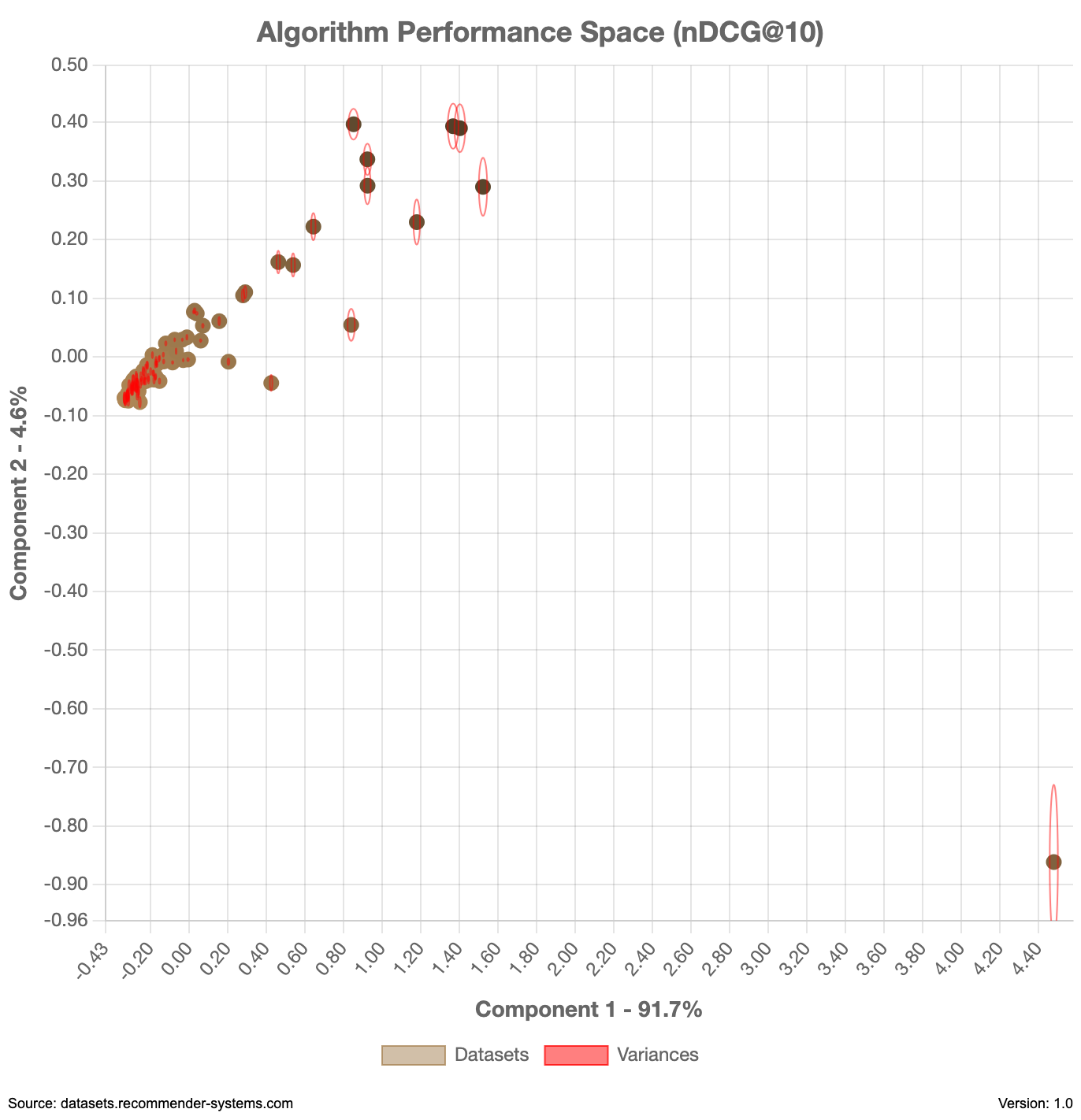}
    \caption[Example of \acl{APS}]{Example of \ac{APS} showing all algorithm with \acs{nDCG}@10}
    \label{fig:aps}
\end{figure}

As it can be seen from \cref{fig:aps}, when constructing the \ac{APS} plot, it was observed that many datasets, particularly those from the Amazon domain, cluster tightly together, indicating low diversity in algorithmic performance. In contrast, datasets like Docear or MovieLens showed more dispersion, suggesting greater utility for comparative evaluation. Notably, this result diverged from findings based on dataset characteristics alone \cite{Chin22}, where Amazon datasets appeared diverse. This discrepancy underscores the central claim of \ac{APS} that performance-based diversity may not align with feature-based or structural dataset diversity, and that the former is more relevant for algorithm evaluation. Furthermore, \acf{PCA} is used to reduce the high-dimensional \ac{APS} into two dimensions. So that the graph becomes easy to read.

In summary, \ac{APS} redefines dataset diversity in practical, empirical terms: by how differently algorithms perform, not by how datasets look on paper. This framework allows for nuanced and rational dataset selection, aimed at improving the validity and generalizability of algorithm evaluations. Importantly, it also provides an extensible foundation. As new algorithms or datasets emerge, they can be embedded into the existing \ac{APS}, making it a living resource that supports reproducible and transparent experimentation. By emphasizing performance-based diversity and enabling empirical justification of dataset selection, \ac{APS} contributes a valuable tool to the recommender-systems community's ongoing efforts toward methodological rigor and scientific progress.

\subsection{Mathematical Background}
\label{subsec:methodology}

This subsection introduces the mathematical methods used in this work. \ac{PCA} is applied to reduce the dimensionality of the \acf{APS}. Quintile-based statistical classification is used to categorize datasets. A variance-normalized distance metric based on the Mahalanobis distance measures dataset similarity and dissimilarity. Finally, an exponential decay function converts distances into similarity confidence scores.”

\subsubsection{Principal Component Analysis}

To better interpret algorithmic performance across a diverse collection of datasets, the application of \acf{PCA} provides a practical method for dimensionality reduction. The goal of this approach is to project high-dimensional performance data derived from metrics such as \ac{nDCG}, \acf{HR}, or Recall at various K-values into a lower-dimensional space that preserves the most salient variance patterns. This transformation allows for more intuitive visual inspection and interpretation of how datasets compare in terms of algorithm performance.

\ac{PCA} generates new variables, referred to as principal components, which are linear combinations of the original performance scores. These components are constructed to be orthogonal to one another and are ordered by the amount of variance they explain. The first principal component (Component 1) captures the direction of greatest variability in the dataset, essentially summarizing the dominant trend in performance across all algorithms and datasets. The second component (Component 2) captures the next highest amount of variance in a direction independent of the first, typically reflecting residual patterns or deviations not accounted for by Component 1.

When plotted in two dimensions using these components as axes, each point represents a dataset, and its position is determined by how it aligns with the underlying performance trends captured by the components. For instance, datasets positioned far apart in this two-dimensional space are those whose performance profiles diverge significantly across the evaluated algorithms and metrics. Conversely, datasets that cluster together indicate similar algorithmic behavior, suggesting potential redundancy in evaluation or similarity in task complexity.

This approach serves multiple purposes. First, it simplifies complex evaluation results into a manageable visualization that facilitates comparisons among datasets. Second, it supports the identification of outlier datasets that deviate substantially from dominant trends, which may be of particular interest for further algorithm development or specialized testing. Third, \ac{PCA} can help to reveal implicit structure in the evaluation results, such as latent performance clusters or dimensions of algorithm sensitivity that may not be visible through raw metric inspection.

Ultimately, the \ac{PCA} offers a powerful lens through which to explore and interpret performance spaces. By reducing dimensionality while preserving the most meaningful variation, it becomes possible to capture the essence of algorithm-dataset interactions and make more informed decisions about dataset selection, algorithm benchmarking, or generalization potential.

\subsubsection{Quintile-Based Statistical Classification}
\label{subsec:quintClass}

A quintile-based classification model divides a dataset into five equal groups based on the distribution of a continuous variable~\cite{Tukey77}. Given a set of values $S = \{s_1, s_2, \ldots, s_n\}$, the method calculates four threshold points that partition the data into quintiles (20\% percentile intervals).
The process begins by sorting all values in ascending order: $s_{(1)} \leq s_{(2)} \leq \ldots \leq s_{(n)}$, and computing the 20th, 40th, 60th, and 80th percentiles following standard statistical procedures~\cite{Casella02}. For a given percentile $p$ (where $0 < p < 100$), the percentile value is calculated using linear interpolation:

\begin{equation*}
\text{percentile}(S, p) = s_{(k)} + (s_{(k+1)} - s_{(k)}) \cdot f
\end{equation*}

where:
\begin{itemize}
   \item $S$ = the set of all values in the dataset
   \item $p$ = the desired percentile (e.g., 20 for the 20th percentile)
   \item $n$ = total number of values in the dataset
   \item $s_{(k)}$ = the $k$-th smallest value in the sorted dataset
   \item $s_{(k+1)}$ = the $(k+1)$-th smallest value in the sorted dataset
   \item $h = \frac{p}{100} \cdot (n-1) + 1$ (interpolation position)
   \item $k = \lfloor h \rfloor$ (the integer part of $h$, representing the lower index)
   \item $f = h - k$ (the fractional part of $h$, used for interpolation between two values)
\end{itemize}

This yields the quintile boundaries:
\begin{align*}
Q_{20} &= \text{percentile}(S, 20) \quad \text{(separates bottom 20\% from rest)} \\
Q_{40} &= \text{percentile}(S, 40) \quad \text{(separates bottom 40\% from top 60\%)} \\
Q_{60} &= \text{percentile}(S, 60) \quad \text{(separates bottom 60\% from top 40\%)} \\
Q_{80} &= \text{percentile}(S, 80) \quad \text{(separates bottom 80\% from top 20\%)}
\end{align*}

These percentiles serve as boundary points to create five distinct groups with equal population:

\begin{itemize}
   \item \textbf{Group 1:} $s \leq Q_{20}$ (0th--20th percentile) - contains 20\% of data
   \item \textbf{Group 2:} $Q_{20} < s \leq Q_{40}$ (20th--40th percentile) - contains 20\% of data
   \item \textbf{Group 3:} $Q_{40} < s \leq Q_{60}$ (40th--60th percentile) - contains 20\% of data
   \item \textbf{Group 4:} $Q_{60} < s \leq Q_{80}$ (60th--80th percentile) - contains 20\% of data
   \item \textbf{Group 5:} $s > Q_{80}$ (80th--100th percentile) - contains 20\% of data
\end{itemize}

This classification method ensures that each group contains exactly 20\% of the observations, providing balanced representation across categories~\cite{Agresti18}. The percentile-based boundaries automatically adapt to the actual distribution of the data, eliminating the need for arbitrary threshold selection that can introduce bias~\cite{Hand01}. The method is robust to outliers since percentiles are less sensitive to extreme values compared to mean-based measures~\cite{Wasserman04}.

Furthermore, the quintile approach provides distributional information beyond simple classification through a relative position measure:

\begin{equation*}
\text{Relative Position}(s_i) = \frac{\text{rank}(s_i) - 1}{n - 1}
\end{equation*}

where:
\begin{itemize}
   \item $s_i$ = a specific value in the dataset
   \item $\text{rank}(s_i)$ = the position of value $s_i$ when all values are sorted from smallest to largest (1 for smallest, $n$ for largest)
   \item $n$ = total number of values in the dataset
   \item Relative Position = a normalized measure between 0 and 1, where 0 represents the smallest value and 1 represents the largest value
\end{itemize}

\subsubsection{Variance-Normalized Distance Metric Based on The Mahalanobis Distance}
\label{sec:variance-distance}
The variance-normalized distance metric is a statistical approach for measuring the similarity between datasets that accounts for the uncertainty in their positions within the \ac{PCA} space. Unlike simple Euclidean distance, which treats all positions as equally certain, this method considers the variance around each dataset's position. The approach is based on the Mahalanobis distance, a well-established statistical measure that adjusts distances according to the variability of the data~\cite{Mahalanobis36}, but simplified for our two-dimensional \ac{PCA} coordinates.

In our implementation, each dataset has $X$ and $Y$ coordinates in the \ac{PCA} space along with corresponding variance values (ellipseX and ellipseY) that represent the uncertainty in these positions. The variance-normalized distance between two datasets is calculated using the formula:

$$\text{distance}^2 = \frac{(x_2 - x_1)^2}{\text{varX}_1 + \text{varX}_2} + \frac{(y_2 - y_1)^2}{\text{varY}_1 + \text{varY}_2}$$

This formula divides the squared coordinate differences by the combined variances along each axis, effectively normalizing the distance by the uncertainty. Moreover, it provides a confidence-adjusted measure of dataset similarity that is more robust than a simple geometric distance, as it accounts for how reliably each dataset's position is determined based on algorithmic performance consistency. The result is then transformed into a similarity confidence score using an exponential decay function:

$$\text{confidence} = e^{-\text{distance}}$$

where smaller distances yield higher confidence scores approaching 1, and larger distances result in confidence scores approaching 0.

\subsubsection{Exponential Decay Function}
The exponential decay function serves as a transformation mechanism to convert the var\-i\-ance-normalized distance into a meaningful similarity confidence score bounded between 0 and 1. This approach addresses the need to quantify dataset similarity in an intuitive manner where higher similarity corresponds to higher confidence scores. The exponential decay function is related to the Gaussian (RBF) kernel~\cite{Scholkopf02}, a widely used method in machine learning and statistics for measuring similarity and statistical confidence between data points.

The standard Gaussian (RBF) kernel formula is:

$$\text{RBF}(x, x') = \exp\left(-\frac{\|x - x'\|^2}{2\sigma^2}\right)$$

where $\|x - x'\|$ represents the distance between two points and $\sigma$ is a parameter that controls the spread or width of the kernel. In this implementation, a simplified form is used that eliminates the need for the scaling parameter:

$$\text{confidence} = e^{-\text{distance}}$$

This simplified version skips the constant factor $1/(2\sigma^2)$ typically found in Gaussian kernels, but maintains the essential exponential decay behavior that makes it effective for similarity measurement.

The practical implementation in our code applies this transformation directly:

$$\text{confidence} = \text{Math.exp}(-\text{distance})$$

where the distance is the variance-normalized metric previously calculated. This exponential decay function exhibits several desirable properties for similarity measurement.

Since the distance is already computed using a variance-normalized distance metric that accounts for positional uncertainty, there is no need for an additional $\sigma$ parameter to control the spread. The variance normalization effectively serves the same role as the $\sigma$ parameter in traditional Gaussian kernels, automatically adapting the similarity calculation to each dataset's uncertainty level. 

\section{Supporting Dataset Selection in Recommender Systems}
\label{sec:mainPart}

This section introduces the \acf{APS} Explorer, a tool to support dataset selection in recommender systems. 
It is a web application organized into three main tabs. The \emph{\acl{APS}} tab visualizes the overall algorithm performance using \acf{PCA}. The \emph{Algorithm Comparison} tab enables direct comparison of the performance of two algorithms. The \emph{Dataset Comparison} tab offers insights into the characteristics of different datasets. Additional export and sharing capabilities are provided as supporting features.

The section is structured as follows. \crefrange{subsec:mainAPS}{subsec:mainDatasetCompare} describe its three main tabs: the \emph{\acl{APS}}, \emph{Algorithm Comparison}, and \emph{Dataset Comparison}. Finally, \cref{subsec:features} presents the export and sharing capabilities.  

\subsection{Algorithm Performance Space Tab}
\label{subsec:mainAPS}

\acl{APS} explorer is a tool to help you select better, more diverse datasets for testing your recommender systems algorithms. Instead of selecting popular datasets blindly, \ac{APS} maps datasets based on how different algorithms perform on them. By using this tool:

\begin{itemize}
    \item You can identify datasets where algorithms struggle.
    \item You can avoid spending much time on datasets where all algorithms already do well.
    \item You can see which datasets are very similar and might not add much new information for training and benchmarking.
\end{itemize}

\subsubsection{Dataset Difficulty}

Dataset difficulty represents how challenging a dataset is for recommender system algorithms to achieve good performance, measured by how algorithms perform on that dataset relative to others in the collection. Dataset difficulty levels are determined using a quintile statistical model that calculates difficulty from their positions on the \ac{PCA} graph~\cite{Beel24} through a two-step process. 

First, the coordinates of each dataset $X$ and $Y$ in the \ac{PCA} space are normalized to a range $[0,1]$, then combined using the arithmetic mean:

\begin{align*}
\text{norm X} &= \frac{\text{result.x} - \text{min X}}{\text{max X} - \text{min X}} \\
\text{norm Y} &= \frac{\text{result.y} - \text{min Y}}{\text{max Y} - \text{min Y}} \\
\text{difficulty-score} &= \frac{\text{norm X} + \text{norm Y}}{2}
\end{align*}

This creates a single difficulty score representing the average normalized position of each dataset in the \ac{PCA} space, where lower scores indicate better algorithmic performance (easier datasets) and higher scores indicate poorer performance (harder datasets). Building on the \ac{APS} framework~\cite{Beel24}, the methodology is extended by replacing arbitrary fixed difficulty thresholds with a data-driven statistical approach.

Second, these difficulty scores are ranked across all datasets and divided into statistical quintiles using the 20th, 40th, 60th and 80th percentiles~\cite{Tukey77} to create five difficulty levels:

\begin{itemize}
    \item \textbf{Very Hard} (0th--20th percentile): Datasets where algorithms struggle the most
    \item \textbf{Hard} (20th--40th percentile): Datasets with below-average algorithm performance
    \item \textbf{Medium} (40th--60th percentile): Datasets with moderate difficulty levels
    \item \textbf{Easy} (60th--80th percentile): Datasets where multiple algorithms achieve above-average performance
    \item \textbf{Very Easy} (80th--100th percentile): Datasets in which most algorithms perform very well
\end{itemize}

This quintile-based approach ensures that exactly 20\% of datasets fall into each difficulty category~\cite{Agresti18}, making the classification inherently relative to the performance characteristics of the entire dataset collection. Unlike the original \ac{APS} implementation~\cite{Beel24}, the dataset difficulty levels are dynamic and adjust automatically. The ranges change whenever you modify the K-value or performance metric selection. This adaptive nature eliminates the need for arbitrary fixed thresholds~\cite{Hand01} and provides meaningful relative comparisons within the current dataset selection.

\subsubsection{Dataset Similarities}

Dataset similarity quantifies the behavioral relationships between datasets based on algorithm performance patterns in the \ac{APS}. To calculate similarity, a variance-normalized distance metric is employed, based on the Mahalanobis distance that accounts for positional uncertainty in the \ac{PCA}-transformed space. The distance between two datasets is computed as described in \cref{sec:variance-distance}.

Dataset similarity is then derived by applying an exponential decay function
which transforms the distance into a confidence score between 0 and 1. Values approaching 1 indicate high similarity (small distances between datasets), meaning algorithms perform similarly across these datasets, while values approaching 0 indicate low similarity (large distances), suggesting different algorithm performance patterns.

\subsubsection{Dataset Dissimilarities}

Dataset dissimilarity serves as the mathematical complement to dataset similarity, measuring how differently algorithms perform across dataset pairs. Dissimilarity maintains a perfect mathematical connection to similarity through the inverse relationship:
\begin{align*}
\text{dissimilarity} = 1 - \text{similarity} = 1 - e^{-\text{distance}}
\end{align*}
This approach ensures that the two measures are inherently connected, where highly similar datasets (similarity $> 0.9$) automatically correspond to low dissimilarity (dissimilarity $< 0.1$), while datasets with low similarity (similarity $< 0.1$) exhibit high dissimilarity (dissimilarity $> 0.9$). The dissimilarity measure uses the same variance-normalized distance foundation as similarity but interprets the results from the opposite perspective. High dissimilarity values (approaching 1.0) indicate that algorithms behave very differently on the dataset pairs, making them valuable for robustness testing and ensuring algorithm generalization across diverse data characteristics. In contrast, low dissimilarity values suggest that datasets share similar algorithm performance patterns and may provide redundant information for algorithm evaluation purposes.

\subsection{Algorithm Comparison Tab}
\label{subsec:mainAlgoCompare}

As a second feature, the \acf{APS} Explorer provides a direct comparison of two selected recommender system algorithms in the tab called \enquote{Algorithm Comparison}. It shows how the two algorithms perform on 96 datasets (\cref{sec:datasetsMetadata}) based on selected performance metrics\footnote{\url{https://datasets.recommender-systems.com/?tab=compareAlgorithms}}. The tool visualizes a scatter plot, as illustrated in \vref{fig:algorithmComparison}. A brief informative paragraph precedes the user options in the algorithm comparison tab. The following section describes the options and interpretation of the algorithm comparison tab.

\subsubsection{Options}

\begin{figure}
    \centering
    \includegraphics[width=\textwidth]{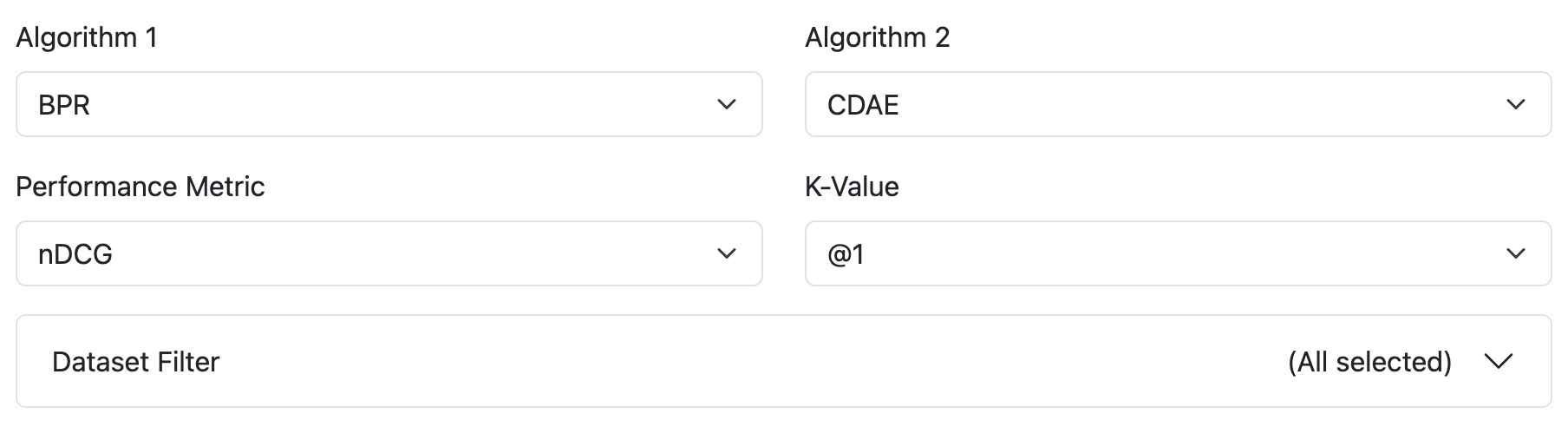}
    \caption[Drop-downs for algorithm comparison selection]{Four drop-downs are available for selecting the options that will form the desired algorithm comparison. Below is an additional drop-down to restrict the datasets used.}
    \label{fig:algorithmComparisonOptions}
\end{figure}

\Cref{fig:algorithmComparisonOptions} shows the interface of the algorithm comparison tab. Every parameter can be changed via a drop-down menu. The user can select two algorithms, a performance metric, a K-value, and the datasets to be included in the comparison. The selected options are then used to generate the scatter plot.

Users can select two different recommender system algorithms and compare their performance across 96 datasets (\cref{sec:datasetsMetadata}). To do so, users first need to select the algorithms from the corresponding drop-downs \enquote{Algorithm 1} and \enquote{Algorithm 2}. The algorithms are sorted alphabetically, and the first two algorithms are selected by default. In the current version, the first two algorithms are \ac{BPR} and \ac{CDAE}. The user can select any two algorithms from the list of 29 available algorithms (including a random recommender) which are listed in (\cref{sec:algorithms}).

The performance metrics applied for analysis can be selected from the drop-down menu labeled \enquote{Performance Metric}. By default, the first option \ac{nDCG} is selected. The available options, which are described in detail in \cref{subsec:metrics}, are:

\begin{itemize} 
    \item \textbf{\ac{nDCG}@K}: Measures ranking quality by rewarding correct recommendations more when they appear higher in the list.
    \item \textbf{\ac{HR}@K}: Checks whether at least one relevant item appears in the recommendation list.
    \item \textbf{Recall@K}: Measures the proportion of all relevant items that appear in the recommendation list.
\end{itemize}

Additionally, users can select the K-value, which determines how many items are considered in the recommendation list. The default value is set to 1, but users can adjust it to any of the values 1, 3, 5, 10, 20.

Afterwards, users can choose the specific datasets they want to include in the comparison. This is directly reflected in the scatter plot, where each point represents a dataset. The datasets can be filtered by selecting the desired checkboxes inside the \enquote{Dataset Filter} drop-down menu. By default, all datasets are selected, but users can choose arbitrary subsets of these datasets. When at least one dataset is not selected, there is also the option to select all datasets via the button \enquote{Select All}. When all datasets are selected, the button changes to \enquote{Deselect All} to allow for quick deselection of all datasets. The datasets are sorted alphabetically in the drop-down menu.

By changing any of the options mentioned above, the scatter plot is automatically updated.

\subsubsection{Interpretation}

The scatter plot visualizes the performance of the two selected algorithms across the chosen datasets. On the x-axis, the performance of algorithm 1 is displayed, while the y-axis shows the performance of algorithm 2. Both axis range from $0$ to $1$, where $0$ indicates the worst performance and $1$ indicates the best performance. This means that each point $x\in[0,1]\times[0,1]$. Each point in the scatter plot represents a dataset, and its x-y-coordinates reflect the performance of both algorithms on that dataset.

The four corners of the coordinate system are colored, to group the data points into a total of five different categories, as can be seen in \vref{fig:algorithmComparison}. These categories are based on the relative performance of the two algorithms on each dataset. The categories, which can also be found in the legend of the scatter plot, are defined as follows:

\begin{itemize}
    \item \textbf{Algorithm 1 outperforms algorithm 2}: The performance of algorithm 1 is better than that of algorithm 2 on the dataset, represented by the blue area in the bottom right corner.
    \item \textbf{Algorithm 2 outperforms algorithm 1}: The performance of algorithm 2 is better than that of algorithm 1 on the dataset, represented by the yellow area in the top left corner.
    \item \textbf{Both algorithms perform well}: Both algorithms achieve the same performance on the dataset, represented by the green area in the top right corner.
    \item \textbf{Both algorithms perform poorly}: Both algorithms achieve poor performance on the dataset, represented by the red area in the bottom left corner.
    \item \textbf{Both algorithms perform moderately}: Both algorithms achieve moderate performance on the dataset, represented by the white area in the center.
\end{itemize}

The colored regions in the scatter plot are intended as a visual aid to help interpret the relative performance of the two algorithms. However, the exact placement of the borders -- such as the division from $(0, 0.5)$ to $(0.5, 1)$ -- is somewhat arbitrary and may be subject to debate. For example, a point at $(0.6, 0)$ appears in the top left corner, but this does not mean that algorithm 2 has solved the problem for that dataset. The borders provide a general orientation but should not be interpreted as strict thresholds.

\begin{figure}
    \centering
    \includegraphics[width=\textwidth]{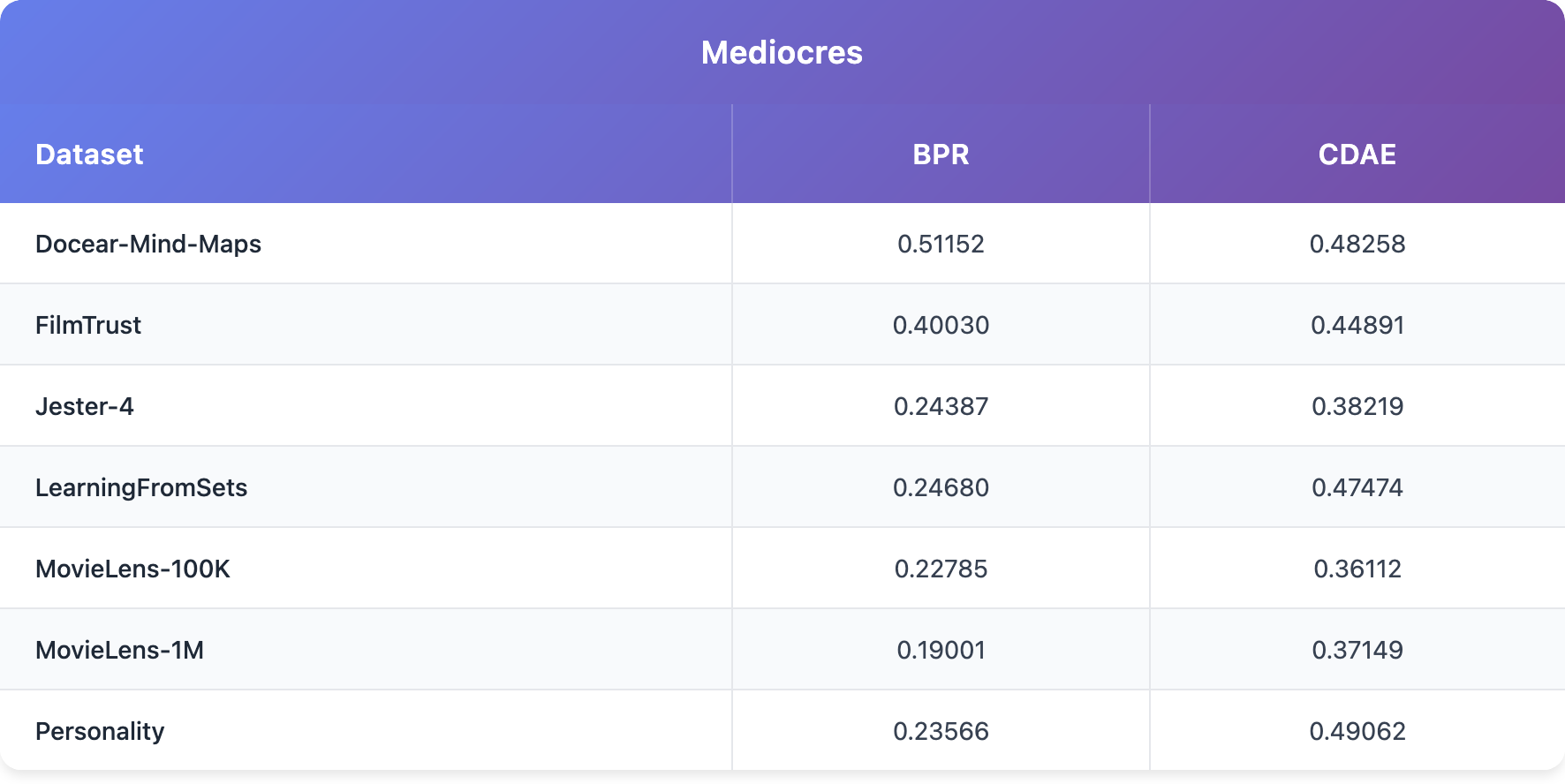}
    \caption[Table of mediocre performance]{The table shows the datasets where both algorithms perform moderately. The performance values of both algorithms are also shown. Selected options are: \ac{BPR} vs. \ac{CDAE} for \ac{nDCG}{@}5}
    \label{fig:algorithmComparisonTable}
\end{figure}

In addition to the scatter plot, the algorithm comparison tab also includes tables that provide a written overview of the aforementioned categories. The tables are located below the scatter plot and show the names of the datasets in each category, along with the performance values of both algorithms. The tables are automatically updated when the scatter plot is changed by selecting different options. As an example the table of mediocre performing algorithms (the white area) is shown in \cref{fig:algorithmComparisonTable}.

\subsubsection{Additional Features}

Finally, there are extra features in the algorithm comparison tab, which are described in detail in \cref{subsec:features}. The features are accessible via dedicated buttons, as can be seen in \vref{fig:algorithmComparison}. It is possible to export the current graph as a PNG image or as a CSV file. Additionally, the user can share the current view by generating a URL that contains the current settings. This allows others to view the same graph without having to manually reapply the filters. The \enquote{See Metadata} button redirects the user to the dataset comparison tab, described in \cref{subsec:mainDatasetCompare}, to see the metadata of the selected datasets.

\subsection{Dataset Comparison Tab}
\label{subsec:mainDatasetCompare}

In this subsection, we will take a look at the Dataset Comparison tab, the third and last tab of the \acf{APS} Explorer. The Dataset Comparison tab of the \ac{APS} Explorer provides additional metadata for all datasets. It is designed to assist researchers in making a final decision about which dataset to use for their experiments. All 96 datasets available in the other tabs of the \ac{APS} Explorer are also accessible in the Dataset Comparison Tab. The user has the flexibility to choose both the datasets to compare and the specific metadata the user wants to view. The datasets and the metadata are visible in a table below the filter of the tab.

The metadata includes various statistics such as the number of users, items, and interactions, user-item ratio, density, mean number of interactions per user and per item, maximum and minimum interactions per user/item, and feedback type (implicit or explicit).
For the columns user-item ratio, mean interactions per item, and mean interactions per user, additional info boxes are provided to offer further context. Columns with additional info boxes are categorized into different risk levels. Since we did not find any research papers that define specific threshold values for risk levels in user-item ratio or mean interactions, we use a quantile-based statistical distribution to determine the risk level ranges. The difference between the quantile-based statistical distribution used for data difficulty or dataset similarity in the \ac{APS} tab and the risk level ranges is that the risk level ranges are static and do not change.

\renewcommand{\arraystretch}{1.5}
\begin{table}
    \centering
    \caption[Risk levels for the user-item]{Risk levels for the user-item ratio using a quantile-based statistical distribution}
    \begin{tabular}{m{0.2\textwidth}m{0.3\textwidth}m{0.3\textwidth}}
        \toprule
        User-item ratio & Description & Cause \\
        \midrule
        $<$ 0.35 & Extremely item-heavy & Too many items with too few users \\
        0.35 - 0.53 & Very item-heavy & Items start getting ignored \\
        0.53 - 1.03 & Slightly item-heavy & Good coverage of items and users \\
        1.03 - 1.66 & Balanced & Optimal learning conditions \\
        1.66 - 2.08 & Slightly user-heavy & Fewer items, users converge on same content \\
        2.08 - 5.16 & Very user-heavy & Too few items for many users \\
        $>$ 5.16 & Extremely user-heavy & Severe lack of items \\
        \bottomrule
    \end{tabular}
    \label{tab:user-item-ratio}
\end{table}

The user-item ratio is an indicator for evaluating the balance of the interaction matrix. An extreme imbalance hinders the model's capacity to learn robust collaborative patterns. In contrast, a more moderate ratio helps adequate interaction overlap between users and items. Therefore, a moderate user-item ratio improves the effectiveness of co-embedding representations. 
The user-item ratio values are divided into seven risk levels. For each level, the potential risk for bias and cold start are indicated, along with possible underlying causes. The risk levels for the user-item ratio are defined in \vref{tab:user-item-ratio}. 

A sufficient volume of user interactions is essential for learning reliable preference patterns in \ac{CF} systems. However, an excessive concentration of interactions among a small subset of highly active users can introduce activity bias, leading the model to overfit to these users and compromising its generalizability. The mean interactions per item and per user are grouped into five risk levels. Similar to the user-item ratio, each level includes interpretation and implications. The risk levels for the mean interactions per user are defined in \vref{tab:mean-interactions-per-user}. 

\begin{table}[ht]
    \centering
    \caption[Risk levels for the mean interactions per user]{Risk levels for the mean interactions per user using a quantile-based statistical distribution}
    \begin{tabular}{m{0.17\textwidth}m{0.5\textwidth}m{0.24\textwidth}}
        \toprule
        Mean interactions per user & Description & Cause \\
        \midrule
        \centering $<$ 8.63 & Low mean interaction per user value with high cold-start risk and long-tail bias & Few users with many interactions \\
        \centering 8.63 - 13.94 & Mostly balanced interaction per users values with some bias and cold-start risk & Some users with many interactions \\
        \centering 13.94 - 20.77 & Balanced mean interaction per user value with low bias and cold-start risk & Balanced distribution of interactions \\
        \centering 20.77 - 28.96 & Few items with a high number of Interactions, which could lead to popularity bias & Some users dominate the interaction space \\
        \centering $>$ 28.96 & Several users with high number of Interactions, which lead to popularity bias & Many users dominate the interaction space \\
        \bottomrule
    \end{tabular}
    \label{tab:mean-interactions-per-user}
\end{table}

Balanced item exposure is crucial for learning robust item representations in recommender systems. Items with sparse interactions tend to be underrepresented in the learned latent space, which results in poor generalization and increase in the cold-start problem. In contrast, items with high interactions can dominate the representation space and can lead to popularity bias. The ratio of the mean interactions per item is divided into five risk levels, as described in \vref{tab:mean-interactions-per-item}. 

\begin{table}[ht]
    \centering
    \caption[Risk levels for the mean interactions per item]{Risk levels for the mean interactions per item using a quantile-based statistical distribution}
    \begin{tabular}{m{0.17\textwidth}m{0.5\textwidth}m{0.24\textwidth}}
        \toprule
        Mean interactions per item & Description & Cause \\
        \midrule
        \centering $<$ 10.48 & Low mean interaction per item value with high cold-start risk and long-tail bias & Few items with many interactions \\
        \centering 10.48 - 15.09 & Mostly balanced Interaction per Item values with some bias and cold-start risk & Some items with many interactions \\
        \centering 15.09 - 22.01 & Balanced Mean Interaction per Item value with low bias and cold-start risk & Balanced distribution of interactions \\
        \centering 22.01 - 66.22 & Few items with a high number of interactions, which could lead to Popularity Bias & Some items dominate the interaction space \\
        \centering $>$ 66.22 & Several items with high number of interactions, which lead to popularity bias & Many items dominate the interaction space \\
        \bottomrule
    \end{tabular}
    \label{tab:mean-interactions-per-item}
\end{table}
\renewcommand{\arraystretch}{1}

An important detail is that all datasets in the \ac{APS} Explorer were subject to 5-core pruning. This means that all users and items with fewer than five interactions were removed. As a result, many datasets have a minimum of five interactions per user or item.
Color-coded indicators in the info boxes are avoided deliberately. While higher risk levels may indicate an increased cold-start risk, such datasets might still be desirable for specific research purposes. A color-coded representation could therefore lead to misunderstandings rather than offering guidance.

\subsection{Export and Sharing Capabilities}
\label{subsec:features}

The website includes several auxiliary features to enhance usability, reproducibility, and collaboration. These features are intended to support researchers in visualizing, analyzing, and sharing experimental outcomes more efficiently. The goal is to offer not only a high-level overview of algorithmic performance but also practical tools for interaction and dissemination. The following features are available in the \acl{APS} tab and the Algorithm Comparison tab:

\subsubsection{Export as PNG}
To facilitate the inclusion of visual outputs in presentations or publications, the interface includes an option to export the current graph as a PNG image. This feature captures the current view of the graph including all applied filters and settings and allows the user to download it with a single click. This ensures that visual insights can be preserved and reused in offline contexts or research papers without requiring external screenshot tools or additional formatting.

\subsubsection{Export as CSV}
For more detailed and structured analysis, the platform provides the option to export graph data as a CSV file. This export includes the full underlying dataset used to generate the current view. This feature allows researchers to conduct their own statistical analyses, build custom visualizations, or integrate \ac{APS} results into larger experimental workflows.

\subsubsection{Share}
The graphs on the website support filtering by dataset and algorithm, allowing users to tailor the visual output to specific needs. Once filters are applied, the share feature generates a unique URL that preserves the current state of the graph. This link can be sent to collaborators or included in supplementary materials, enabling others to view the exact same graph configuration without manually reproducing the filtering steps. As a result, this functionality promotes reproducibility and simplifies collaborative review of experiment results.

\subsubsection{See Metadata}
The metadata button provides a quick redirection from the current tab to the dataset comparison tab. When the dataset filter is applied to the current graph, clicking this button redirects the user to the dataset comparison tab while preserving the same dataset filter. This eliminates the need to manually reapply filters on the metadata view and ensures a seamless transition between performance visualization and contextual dataset information.

\section{Implementation}
\label{sec:implementation}

This section provides an overview of the implementation details of the \acf{APS} Explorer web application. It covers the technologies used, the project structure, and the key components that make up the application.

\subsection{System Architecture Overview}

The application follows a standard web architecture with clear separation between front end and back end components. The front end consists of static HTML files, JavaScript modules for dynamic functionality, and CSS styling with Bootstrap. The back end implements a RESTful API in PHP that interfaces with the MySQL database.

The front end implements a \ac{SPA} using vanilla JavaScript with ES6 modules, structured around three main components: \ac{APS} for \acf{PCA} visualization, Algorithm Comparison for performance analysis, and Dataset Comparison for tabular data presentation. The system uses Bootstrap 5 for responsive design and Chart.js for interactive visualizations, with specialized modules like \texttt{chartHelper.js} providing abstraction layers. Dynamic content loading through \texttt{dynamicContent.js} enables efficient resource utilization by fetching HTML templates on-demand and initializing corresponding JavaScript modules.

The back end implements a RESTful \ac{API} in PHP with MySQL, using centralized routing and modular handlers for different entities. Security is enforced via prepared statements and header-based authentication. Front end - back end integration is managed by a centralized \texttt{ApiService} with caching and error handling. The system supports both client-side and server-side \ac{PCA} calculations for performance and responsiveness.

The system features a modular chart rendering pipeline, URL-based state management, and responsive design. JavaScript modules separate data fetching, processing, and presentation.

\subsection{Technology Stack}

The project is built using a common combination of web technologies that include HTML and JavaScript for the front end and PHP for the back end. The design of the front end is enhanced with the Bootstrap library.

\subsubsection{Front End Technologies}

\begin{lstlisting}[float, language=HTML, caption={HTML Structure for \ac{APS} Explorer}, label=lst:html-aps-explorer, firstnumber=70]
<div class="container" style="max-width: 800px">
  <ul class="nav nav-tabs mb-3" id="tab" role="tablist">
    <li class="nav-item" role="presentation">
      <button id="aps-tab-btn" class="nav-link active" id="aps-tab" data-bs-toggle="tab" data-bs-target="#aps-tab-pane" type="button" role="tab"
        aria-controls="aps-tab-pane" aria-selected="true">
        Algorithm Performance Space
      </button>
    </li>
    [...]
  </ul>
  <div class="container">
    <div class="tab-content" id="tabContent">
      <!-- Dynamic content -->
    </div>
    <div class="loading" id="loadingContainer" style="display: flex; flex-direction: column; justify-content: start;">
      <i class="spinner-border text-light loading-icon" style="margin-top: 5rem;"></i>
    </div>
  </div>
</div>
\end{lstlisting}

HTML serves as the backbone of the user interface, providing a semantic structure for the application. It ensures that the content is accessible and well-organized, allowing users to navigate through different sections easily. \cref{lst:html-aps-explorer}, shows an example of the HTML structure of the tab system for the \ac{APS} Explorer, which is a key component of the application.

The outer \texttt{div} provides the container for the entire tab system. Inside this container, an unordered list (\texttt{ul}) is used to define the navigation tabs. Each tab is represented by a list item (\texttt{li}) containing a button that activates the corresponding tab pane when clicked. The actual content of the tab is dynamically loaded into the \texttt{div} with the class \texttt{tab-content}. The loading indicator is also included in the HTML structure, which is displayed while the content is being fetched.

JavaScript is used to enable dynamic interactions within the application. The three tabs for example, are loaded dynamically using JavaScript, allowing users to switch between different views without reloading the page. This enhances the user experience by providing a seamless and responsive interface. The implementation of dynamic tab loading is shown in \cref{lst:js-tab-loading}, where the JavaScript class \texttt{DynamicContent} is responsible for fetching and injecting HTML content into the specified parent element.

\begin{lstlisting}[float, language=JavaScript, caption={JavaScript for Dynamic Tab Loading}, label=lst:js-tab-loading]
export class DynamicContent {
    static async loadContentAndFindElementById(htmlFileName, parentElementId) {
        const file = await fetch(htmlFileName);
        const element = document.getElementById(parentElementId);
        element.innerHTML = await file.text();
        return element;
    }
    static async loadContentToElement(htmlFileName, parentElement) {
        const file = await fetch(htmlFileName);
        parentElement.innerHTML = await file.text();
    }
}
\end{lstlisting}

Bootstrap\footnote{https://getbootstrap.com/} is used to style the user interface. This is accomplished by applying Bootstrap classes to HTML elements, which provides a consistent and visually appealing design. This can be seen for example in \cref{lst:html-aps-explorer}, where Bootstrap classes such as \texttt{container}, \texttt{nav-tabs}, and \texttt{spinner-border} are used to create a responsive layout and loading indicator.

To visualize the graphs in the tabs \emph{\acl{APS}} and \emph{Algorithm Comparison} the charting library Chart.js\footnote{https://www.chartjs.org/} is employed. It enables the creation of interactive and responsive charts with minimal effort. The library supports various chart types, including line, bar, and pie charts. The \ac{APS} Explorer only uses scatter charts for visualizations. Chart.js is lightweight and easy to integrate, making it an ideal choice for our project.

\subsubsection{Back End Technologies}

The back end uses PHP for server-side processing and MySQL for data storage. PHP handles HTTP requests and returns data in JSON format. MySQL stores the algorithms, datasets, performance results, and precomputed \ac{PCA} values in four main tables.

\subsubsection{Data Processing Libraries}

\ac{PCA} is handled both server-side and client-side depending on the presence of user-applied filters. When no filters are selected, meaning all datasets and algorithms are included, the system fetches precomputed \ac{PCA} results directly from the back end. These results are stored in the database and rendered immediately in the interface, enabling fast, zero-computation loading of the full \ac{PCA} graph.

When filters are applied, \ac{PCA} is recomputed on the client. In this case, the system fetches only the performance results corresponding to the selected datasets and algorithms. The \ac{PCA} is then dynamically calculated in the browser using a JavaScript \ac{PCA} library called ML-\ac{PCA}\footnote{https://www.skypack.dev/view/ml-pca}. To minimize redundant requests and optimize performance, the system maintains an internal cache of previously fetched results. Any new filter operation triggers additional fetches only for missing data, ensuring the \ac{PCA} is always computed with the most complete available matrix. This hybrid approach balances responsiveness and scalability, particularly when analyzing large sets of combinations with interactive filtering.

\subsection{Database Design}
\label{sec:database}

For a database, MySQL is used to support persistent storage of all necessary entities and results. The database schema includes the following four tables.

\begin{itemize}
\item \textbf{Algorithms}: Stores all algorithms' names
\item \textbf{Datasets}: Stores all datasets with their metadata.
\item \textbf{PerformanceResults}: Stores all the performance of each algorithm-dataset pair for each metric (\ac{nDCG}, \acf{HR}, Recall) and K-values (1, 3, 5, 10, 20).
\item \textbf{PcaResults}: Stores precomputed \ac{PCA} values for all datasets in the full \ac{APS} graph.
\end{itemize}

This schema allows efficient querying and aggregation while supporting both static (precomputed) and dynamic (client-computed) analysis pipelines.

\subsection{Back end Implementation}
\label{sec:backend}

The back end is implemented in PHP and acts as the \ac{API} layer between the client and the database. The \texttt{index.php} entry point routes all incoming requests based on an action parameter and delegates operations to the appropriate handlers. Public endpoints include functions for retrieving algorithms, datasets, performance results and \ac{PCA} data. All responses are returned in JSON format and have appropriate HTTP status codes to handle errors such as missing parameters or invalid IDs.

The back end performs validation for all input parameters to ensure type safety and prevent misuse. For instance, all IDs must be integers and passed as arrays where required. Modular API files manage distinct entities such as datasets, algorithms, and performance results. This allows for separation of concerns and easier maintenance.

\subsubsection{Admin APIs}

Administrative functions are handled by a dedicated set of endpoints under the admin action. These endpoints include the following features.

\begin{itemize}
\item Adding new performance results
\item Retrieving raw performance data
\item Updating stored \ac{PCA} results
\end{itemize}

All administrative \acp{API} are protected via a header-based authentication mechanism. A valid admin key and secret must be provided with each request.

In addition, a kill switch mechanism is added on the server to prevent malicious attempts to use admin \acp{API}. All admin endpoints automatically reject requests and return HTTP 403 (Forbidden) if the admin secret key is set to null. This mechanism provides a simple yet effective safeguard for preventing unauthorized or accidental modifications in the database.

\section{Conclusion}
\label{sec:conclusion}

In this report, the \acf{APS} Explorer was introduced, a web-based visualization tool designed to support the dataset selection for recommender systems. The \ac{APS} Explorer aims to address the issue of researchers selecting datasets for recommender system experiments without any justification. The \ac{APS} Explorer offers visualization of the performance of 29 algorithms on 96 datasets on metrics such as \ac{nDCG}, \ac{HR} and Precision. The core functionality across the three interactive tabs were demonstrated: \emph{\acl{APS}}, \emph{Algorithm Comparison} and \emph{Dataset Comparison}.

The \acl{APS} tab provides an overview of how different algorithms perform across multiple datasets, helping users to spot general trends and outliers. The \acl{APS} uses dimensionality reduction techniques like \acf{PCA} to project performance of multiple algorithms across datasets into a 2D space. The \ac{APS} tab also provides information about dataset difficulty and dataset similarities. The algorithm comparison tab shows how two algorithms perform across a set of datasets, highlighting where one outperforms the other, where both succeed, or where both struggle. The dataset comparison tab enables users to evaluate which datasets are more or less challenging for recommendation tasks, based on the number of users, items and interactions. Risk levels were assigned to datasets based on interaction density and the user-item ratio, helping to identify potential sources of bias or cold-start problems. These components enable users to explore how different algorithms perform across various datasets and help them identify suitable combinations for their specific use cases.

Looking ahead, there are several promising directions to expand and improve the \ac{APS} Explorer. One important enhancement would be the integration of additional datasets and algorithms, offering a broader exploration space and increasing the tool's applicability. The integration of additional datasets and algorithms can be accomplished by adding a function in the \ac{APS} Explorer for adding new datasets and algorithms. To further support researchers in interpreting the evaluation results, enhanced visual explanations and features could be added to the interface. Moreover, incorporating user studies would provide valuable feedback on the usability and practical value of the tool, helping to refine the design of the \ac{APS} explorer. 

\section{Acknowledgments}

We acknowledge the use of ChatGPT to refine the clarity and flow of our writing, and DeepL to assist with accurate translation. Both tools were employed solely as aids for grammar, phrasing, and formulation.

\printbibliography[title=Bibliography]

\newpage
\addsec{List of Abbreviations}
\begin{acronym}[nDCG]  		
\acro{APS}{Algorithm Performance Space}
\acro{API}{application programming interface}
\acro{BPR}{Bayesian Personalized Ranking}
\acro{CDAE}{Collaborative Denoising Auto-Encoders}
\acro{CF}{collaborative filtering}
\acro{HR}{hit ratio}
\acro{nDCG}{normalized discounted cumulative gain}
\acro{PCA}{principal component analysis}
\acroplural{PCA}[PCAs]{principal component analyses}
\acro{PDO}{PHP Data Objects}
\acro{SPA}{Single Page Application}
\end{acronym}

\listoftables

\lstlistoflistings

\listoffigures

\appendix

\newgeometry{margin=1cm}
\pagestyle{empty}
\section{Metadata of Datasets}
\label{sec:datasetsMetadata}
\scriptsize
\rowcolors{1}{}{black!10!white}
\begin{longtable}{lllll}
\toprule
Dataset & Number of Users & Number of Items & Number of Interactions & User-Item Ratio \\
\midrule
\endfirsthead

\toprule
Dataset & Number of Users & Number of Items & Number of Interactions & User-Item Ratio \\
\midrule
\endhead

\midrule
\multicolumn{5}{r}{{Continued on next page}} \\
\midrule
\endfoot

\bottomrule
\endlastfoot

Amazon-Ratings & 99168 & 106535 & 1675292 & 0.93 \\
Amazon2014-Amazon-Instant-Video & 5130 & 1685 & 37126 & 3.04 \\
Amazon2014-Apps-For-Android & 87271 & 13209 & 752937 & 6.61 \\
Amazon2014-Automotive & 2928 & 1835 & 20473 & 1.60 \\
Amazon2014-Baby & 19445 & 7050 & 160792 & 2.76 \\
Amazon2014-Beauty & 22363 & 12101 & 198502 & 1.85 \\
Amazon2014-CDs-And-Vinyl & 75258 & 64443 & 1097592 & 1.17 \\
Amazon2014-Cell-Phones-And-Accessories & 27879 & 10429 & 194439 & 2.67 \\
Amazon2014-Clothing-Shoes-And-Jewelry & 39387 & 23033 & 278677 & 1.71 \\
Amazon2014-Digital-Music & 5541 & 3568 & 64706 & 1.55 \\
Amazon2014-Grocery-And-Gourmet & 14681 & 8713 & 151254 & 1.68 \\
Amazon2014-Health-And-Personal-Care & 38609 & 18534 & 346355 & 2.08 \\
Amazon2014-Home-And-Kitchen & 66519 & 28237 & 551682 & 2.36 \\
Amazon2014-Kindle-Store & 68223 & 61934 & 982619 & 1.10 \\
Amazon2014-Musical-Instruments & 1429 & 900 & 10261 & 1.59 \\
Amazon2014-Office-Products & 4905 & 2420 & 53258 & 2.03 \\
Amazon2014-Patio-Lawn-And-Garden & 1686 & 962 & 13272 & 1.75 \\
Amazon2014-Pet-Supplies & 19856 & 8510 & 157836 & 2.33 \\
Amazon2014-Sports-And-Outdoors & 35598 & 18357 & 296337 & 1.94 \\
Amazon2014-Tools-And-Home-Improvement & 16638 & 10217 & 134476 & 1.63 \\
Amazon2014-Toys-And-Games & 19412 & 11924 & 167597 & 1.63 \\
Amazon2014-Video-Games & 24303 & 10672 & 231780 & 2.28 \\
Amazon2018-Arts-Crafts-And-Sewing & 21050 & 45346 & 392935 & 0.46 \\
Amazon2018-Automotive & 76701 & 179281 & 1573795 & 0.43 \\
Amazon2018-Cell-Phones-And-Accessories & 47284 & 153415 & 1098877 & 0.31 \\
Amazon2018-Clothing-Shoes-And-Jewelry & 24584 & 59119 & 412969 & 0.42 \\
Amazon2018-Digital-Music & 9906 & 12381 & 123518 & 0.80 \\
Amazon2018-Fashion & 7 & 374 & 2588 & 0.02 \\
Amazon2018-Gift-Cards & 147 & 453 & 2941 & 0.32 \\
Amazon2018-Grocery-And-Gourmet-Food & 39504 & 113592 & 1005817 & 0.35 \\
Amazon2018-Industrial-And-Scientific & 4353 & 8342 & 58664 & 0.52 \\
Amazon2018-Luxury-Beauty & 936 & 2028 & 20050 & 0.46 \\
Amazon2018-Magazine-Subscriptions & 132 & 294 & 2027 & 0.45 \\
Amazon2018-Musical-Instruments & 9930 & 24780 & 206241 & 0.40 \\
Amazon2018-Office-Products & 25898 & 86713 & 677247 & 0.30 \\
Amazon2018-Patio-Lawn-And-Garden & 30761 & 89424 & 685274 & 0.34 \\
Amazon2018-Prime-Pantry & 4898 & 13073 & 126702 & 0.37 \\
Amazon2018-Software & 643 & 1339 & 9596 & 0.48 \\
Amazon2018-Sports-And-Outdoors & 69781 & 185024 & 1498609 & 0.38 \\
Amazon2018-Video-Games & 16882 & 50626 & 453881 & 0.33 \\
Behance & 23724 & 29794 & 687070 & 0.80 \\
BookCrossing & 7025 & 9432 & 118668 & 0.74 \\
CiaoDVD & 1807 & 2062 & 28059 & 0.88 \\
CiteULike-A & 5536 & 15429 & 200180 & 0.36 \\
CiteULike-T & 3527 & 6339 & 77546 & 0.56 \\
DeliveryHero-SE & 36088 & 17932 & 347987 & 2.01 \\
Diginetica & 6459 & 3905 & 46115 & 1.65 \\
Docear-Mind-Maps & 1769 & 578 & 17271 & 3.06 \\
Douban-Book & 25477 & 30484 & 1110912 & 0.84 \\
Douban-Music & 18604 & 29981 & 1041460 & 0.62 \\
Douban-Short & 108432 & 28 & 961228 & 3872.57 \\
Epinions & 20397 & 21901 & 446892 & 0.93 \\
Escorts & 1474 & 1021 & 15887 & 1.44 \\
FilmTrust & 1208 & 406 & 31668 & 2.98 \\
FoodComRecipes & 17813 & 41240 & 555618 & 0.43 \\
Foursquare-NYC1 & 547 & 547 & 4524 & 1.00 \\
Foursquare-NYC2 & 1062 & 3895 & 40825 & 0.27 \\
Foursquare-Tokyo & 2287 & 7055 & 128530 & 0.32 \\
Frappe & 576 & 550 & 13116 & 1.05 \\
GoogleLocal2018 & 57395 & 71434 & 893214 & 0.80 \\
GoogleLocal2021-Alaska & 42382 & 8738 & 678297 & 4.85 \\
GoogleLocal2021-Delaware & 70498 & 11089 & 1146255 & 6.36 \\
GoogleLocal2021-District-Of-Columbia & 73064 & 7711 & 870531 & 9.48 \\
GoogleLocal2021-New-Hampshire & 101978 & 18228 & 1655444 & 5.59 \\
GoogleLocal2021-North-Dakota & 43536 & 8655 & 714773 & 5.03 \\
GoogleLocal2021-Rhode-Island & 66629 & 12096 & 1117324 & 5.51 \\
GoogleLocal2021-South-Dakota & 57728 & 10291 & 890388 & 5.61 \\
GoogleLocal2021-Vermont & 32987 & 7604 & 459769 & 4.34 \\
GoogleLocal2021-Wyoming & 44081 & 8254 & 606353 & 5.34 \\
Hetrec-LastFM & 1090 & 3646 & 52551 & 0.30 \\
IPinYou & 1467 & 93 & 7857 & 15.77 \\
Jester-4 & 4096 & 136 & 97879 & 30.12 \\
KGRec-Music & 5199 & 8640 & 751531 & 0.60 \\
LearningFromSets & 854 & 8870 & 450123 & 0.10 \\
Librarything & 23532 & 47415 & 827491 & 0.50 \\
MarketBias-ModCloth & 763 & 2628 & 39652 & 0.29 \\
ModCloth-Clothing-Fit & 2238 & 301 & 15798 & 7.44 \\
MovieLens-100K & 943 & 1349 & 99287 & 0.70 \\
MovieLens-1M & 6040 & 3416 & 999611 & 1.77 \\
MovieLens-Latest-Small & 610 & 3650 & 90274 & 0.17 \\
MovieTweetings & 23421 & 11888 & 802784 & 1.97 \\
Myket-Android & 9983 & 7814 & 545591 & 1.28 \\
Personality & 1819 & 14868 & 984499 & 0.12 \\
Rekko & 24523 & 4305 & 291882 & 5.70 \\
RentTheRunway & 4911 & 3166 & 42380 & 1.55 \\
Retailrocket & 22178 & 17803 & 240938 & 1.25 \\
Sketchfab & 25587 & 15254 & 546349 & 1.68 \\
StackOverflow & 34151 & 23098 & 470966 & 1.48 \\
Steam-Australian-Reviews & 1761 & 443 & 10730 & 3.98 \\
TaFeng & 26039 & 15483 & 709356 & 1.68 \\
Twitch-Full & 2067 & 967 & 11267 & 2.14 \\
WikiLens & 241 & 1594 & 20539 & 0.15 \\
Yahoo-Movies & 7620 & 3783 & 207771 & 2.01 \\
Yahoo-Music3 & 15400 & 1000 & 365704 & 15.40 \\
Yahoo-Semi-Synthetic & 7597 & 3675 & 197095 & 2.07 \\
Yoochoose & 150996 & 16919 & 1014173 & 8.92 \\

\end{longtable}

\normalsize
\newpage
\pagestyle{scrheadings}
\restoregeometry
\section{List of Algorithms}
\label{sec:algorithms}

The following 28 algorithms were used in our experiments \cite{RecBoleModels}:

\begin{multicols}{3}
\begin{itemize}
\item BPR
\item CDAE
\item DGCF
\item DMF
\item ENMF
\item ItemKNN
\item LightGCN
\item LINE
\item MultiDAE
\item MultiVAE
\item NeuMF
\item NGCF
\item Pop
\item RecVAE
\item SGL
\item SimpleX
\item SpectralCF
\item ConvNCF
\item DiffRec
\item EASE
\item FISM
\item GCMC
\item LDiffRec
\item MacridVAE
\item NAIS
\item NCEPLRec
\item NCL
\item NNCF
\end{itemize}
\end{multicols}

\end{document}